\documentclass[aps,prd,twocolumn,showpacs,floatfix,preprintnumbers,amsfont,amsmath,amssymb,nofootinbib, superscriptaddress, reprint]{revtex4-2}

\usepackage[usenames,dvipsnames,svgnames,table]{xcolor} 
\usepackage{amsmath, amssymb, amsbsy, amstext, amsthm, simplewick}
\usepackage{hyperref}
\usepackage{amsfonts}
\usepackage{latexsym, mathrsfs}
\usepackage{graphicx}

\makeatletter
\newsavebox\myboxA
\newsavebox\myboxB
\newlength\mylenA
\newcommand*\xoverline[2][0.65]{%
    \sbox{\myboxA}{$\m@th#2$}%
    \setbox\myboxB\null
    \ht\myboxB=\ht\myboxA%
    \dp\myboxB=\dp\myboxA%
    \wd\myboxB=#1\wd\myboxA
    \sbox\myboxB{$\m@th\overline{\copy\myboxB}$}
    \setlength\mylenA{\the\wd\myboxA}
    \addtolength\mylenA{-\the\wd\myboxB}%
    \ifdim\wd\myboxB<\wd\myboxA%
       \rlap{\hskip 0.5\mylenA\usebox\myboxB}{\usebox\myboxA}%
    \else
        \hskip -0.5\mylenA\rlap{\usebox\myboxA}{\hskip 0.5\mylenA\usebox\myboxB}%
    \fi}
\makeatother

\usepackage{xr}
\begin{document}
\title{Tidal Deformation and Dissipation of Rotating Black Holes}
\author{Horng Sheng Chia}
\affiliation{School of Natural Sciences, Institute for Advanced Study, Princeton, NJ 08540, USA}
\affiliation{Institute of Physics, University of Amsterdam, Amsterdam, 1098 XH, the Netherlands}

\definecolor{Red}{rgb}{0.92,0.,0.}
\newcommand{\hs}[1]{\textcolor{red}{(HS: #1)}}
\newcommand{\HS}[1]{\textcolor{red}{#1}}
\def\d{{\rm d}}
\def\hyp{${}_2${\fontfamily{cmss}\selectfont F}$_1$}
\def\beq{\begin{equation}}
\def\eeq{\end{equation}}
\renewcommand{\appendixname}{Appendix}

\begin{abstract}

We show that rotating black holes do not experience any tidal deformation when they are perturbed by a weak and adiabatic gravitational field. The tidal deformability of an object is quantified by the so-called ``Love numbers", which describe the object's linear response to its external tidal field. In this work, we compute the Love numbers of Kerr black holes and find that they vanish identically. We also compute the dissipative part of the black hole's tidal response, which is non-vanishing due to the absorptive nature of the event horizon. Our results hold for arbitrary values of black hole spin, for both the electric-type and magnetic-type perturbations, and to all orders in the multipole expansion of the tidal field. The boundary conditions at the event horizon and at asymptotic infinity are incorporated in our study, as they are crucial for understanding the way in which these tidal effects are mapped onto gravitational-wave observables. In closing, we address the ambiguity issue of Love numbers in General Relativity, which we argue is resolved when those boundary conditions are taken into account. Our findings provide essential inputs for current efforts to probe the nature of compact objects through the gravitational waves emitted by binary systems.

\end{abstract}

\maketitle

\section{Introduction}

An important open problem in astrophysics today is how black holes would respond to the perturbation sourced by an external tidal field. The tidal response of a self-gravitating object consists of a conservative part, which describes how the object would deform, and a dissipative part, which quantifies the amount of energy that would be lost due to this tidal interaction. More precisely, the tidal deformability of an object is characterized by the so-called ``Love numbers"~\cite{Love, poisson_will_2014, MurrayDermott}, which quantify the induced moments the object would acquire when the tidal environment is static. On the other hand, tidal dissipation arises due to the viscosity of the object, and is only present if the environment varies with time in the object's rest frame~\cite{poisson_will_2014, MurrayDermott}. These tidal effects encode information about the object's internal structure, and appear at different post-Newtonian (PN) orders in the phase of the gravitational waves emitted when the object is part of a binary system. More precisely, the tidal deformation of a body first appears at 5PN order in the phase of a binary waveform~\cite{Flanagan:2007ix, Vines:2011ud, Bini:2012gu}, while tidal dissipation of a rotating (non-rotating) body first appears at 2.5PN (4PN) order~\cite{Poisson:1994yf, Tagoshi:1997jy, Alvi:2001mx, Poisson:2004cw, Goldberger:2005cd, Porto:2007qi, Chatziioannou:2012gq}. Gravitational-wave sources are often intrinsically dark -- a precise measurement of these tidal effects would not only offer us valuable probes into the nature of known objects, such as neutron stars~\cite{Hinderer:2007mb, Hinderer:2009ca, Damour:2012yf, TheLIGOScientific:2017qsa, Abbott:2018exr}, but could also provide hints for the existence of new types of compact objects~\cite{Giudice:2016zpa, Sennett:2017etc, Cardoso:2017cfl, Baumann:2018vus, Baumann:2019ztm, Datta:2020gem}. Black holes are the simplest and most fascinating self-gravitating objects~\cite{Kerr1963, Carter:1971zc, Robinson:1975bv}; a detailed understanding of their tidal effects is not only of direct interest in astrophysics, but could also provide further insights into the foundational aspects of black holes and gravity. 

\vskip 2pt

While the Love numbers of Schwarzschild black holes have been shown to vanish~\cite{Binnington:2009bb, Damour:2009vw, Kol:2011vg, Gurlebeck:2015xpa, Hui:2020xxx}, similar conclusions for Kerr black holes have only been drawn in the small-spin limit and for the few low-order multipolar perturbations~\cite{Landry:2015zfa, Pani:2015hfa, Pani:2015nua}. In the latter case, the results are limited because the metric of an arbitrarily spinning black hole in a general tidal environment is hard to derive~\cite{Poisson:2014gka, Pani:2015hfa}. Finding the Love numbers through the metric has also been shown to suffer from a drawback: the meaning of Love numbers seem to depend on coordinate choices and are thus ambiguous~\cite{Fang:2005qq, Pani:2015hfa, Gralla:2017djj}. Recently, Ref.~\cite{LeTiec:2020spy} attempted at computing the Love numbers of the Kerr black hole through the Weyl scalars of the Newman-Penrose formalism~\cite{NP1962}. This approach is advantageous because the technical heft of deriving the perturbed Kerr metric is now replaced by solving the Teukolsky equation~\cite{Teukolsky1972a, Teukolsky1973b}, which is a much simpler task. As we shall see, the resolution to the ambiguity issue described above lies in a careful consideration of the boundary condition at asymptotic infinity, which is also captured by the Teukolsky equation. 

\vskip 2pt

In this work, we compute the tidal deformation and dissipation of rotating black holes through a detailed examination of the $\psi_4$ Weyl scalar~\cite{NP1962} for the tidally-perturbed Kerr black hole. Our work crucially exploits the separability of $\psi_4$ on the Kerr background~\cite{Teukolsky1972a, Teukolsky1973b}, which allows us to solve for the black hole's responses even when the details of the perturbed metric is not fully known. We make no assumption about the spin of the black hole, and only assume that the external tidal field sources a linear perturbation that is slowly varying with time; our results are otherwise true for all spins and to all orders in the multipole expansion of the linear external tidal field. We describe how the black hole's tidal deformation and dissipation directly affect the outgoing flux at asymptotic infinity and ingoing flux at the event horizon, respectively. Since these fluxes are coordinate-invariant observables, the tidal effects that we compute have unambiguous physical interpretations on gravitational waveforms. 

\vskip 2pt

Through the Newman-Penrose formalism, the authors of Ref.~\cite{LeTiec:2020spy} claimed that the tidal deformability of rotating black holes do not vanish for non-axisymmetric perturbations -- a result which is in direct tension with Ref.~\cite{Poisson:2014gka}, which showed that the Love numbers of non-axisymmetric quadrupolar perturbations are zero for a slowly-rotating black hole. In this work, we resolve this issue by showing that so called ``non-vanishing Love numbers" found by Ref.~\cite{LeTiec:2020spy} are not associated to the conservative tidal deformability of the Kerr black hole, but are instead dissipative effects of the black hole. In particular, the ``Love numbers" computed in Ref.~\cite{LeTiec:2020spy}, which are purely imaginary quantities, do not vanish for a static external tidal field because the black hole's rotational motion sources a relative time dependence with the static environment. In other words, in the co-rotating frame of the Kerr black hole, a static tidal field would be perceived to rotate at a frequency that is proportional to the black hole spin, which therefore induces tidal dissipation.

\vskip 4pt

\textit{Notation and conventions.---} We use the $(-, +, +, +)$ metric convention and work in units of $G=c=1$. We adopt the ingoing-Kerr coordinates $\{v, r, \theta,  \phi \}$, in which the line element of the Kerr black hole with mass $M$ and specific angular momentum $a$ is
\beq
\begin{aligned}
\d & s^2 = - (\Delta - a^2 \sin^2 \theta) \hskip 1pt \d v^2 /\Sigma + 2 \hskip 1pt dv \d r \\
& - 2 a \sin^2 \theta \hskip 1pt  \d r \d \phi  - 4 M a r \sin^2 \theta  \hskip 1pt \d v \d \phi / \Sigma    \\
& + \Sigma \hskip 1pt \d \theta^2 + \left[ (r^2 + a^2)^2 - \Delta a^2 \sin^2 \theta \right] \sin^2 \theta  \hskip 1pt \d \phi^2 /  \Sigma \, , \label{eqn:Kerrcoord}
\end{aligned}
\eeq
where $\Delta = (r-r_+)(r-r_-), \Sigma = r^2 + a^2 \cos^2 \theta$, and $r_{\pm} = M \pm (M^2 - a^2)^{1/2}$ are the inner and outer horizons of the black hole. The sum $\sum_{\ell m}$ includes the total angular momentum numbers $\ell \geq 2$ and the azimuthal numbers $|m|\leq \ell$. The index $I=\{E, B\}$ labels quantities of electric and magnetic character respectively, and $\sum_I$ represents the sum over both types of quantities. We follow the normalization convention of Refs.~\cite{Binnington:2009bb, Zhang1986} for the tidal moments and their multipole expansions.

\vskip 4pt
\section{The $\psi_4$\hskip1pt Weyl scalar}
The $\psi_4$ curvature invariant of the Newman-Penrose formalism~\cite{NP1962} describes the two transverse polarizations of the gravitational waves that propagate towards asymptotic infinity~\cite{Szekeres:1965ux}. For a Kerr black hole immersed in a weak gravitational field, the Teukolsky equation~\cite{Teukolsky1972a, Teukolsky1973b} separates $\rho^4 \psi_4$, where $\rho \equiv -(r - i a \cos \theta)$, into a set of coupled angular and radial ordinary differential equations. Denoting the mode frequency and azimuthal number by $\omega$ and $m$, the separable form reads
\beq
\begin{aligned}
\rho^4 \psi_4  =  \sum_{\ell m}  \, e^{- i \omega v + i m \phi}  \hskip 1pt \xoverline{R}_{\ell m}(r) {}_{-2} S_{\ell m} (\theta) \, ,  \label{eqn:mode}
\end{aligned}
\eeq
where $\xoverline{R}$ is the radial function, ${}_{-2} S_{\ell m}$ is the spin-weighted spheroidal harmonic with spin parameter $s=-2$~\cite{Teukolsky1972a, Teukolsky1973b}, and $\ell$ is the total angular momentum number. Importantly, the separable form (\ref{eqn:mode}) is true for \textit{any} linear gravitational perturbation on the Kerr background. The precise nature of the perturbation is encoded in the sizes of the mode amplitudes, which we have implicitly absorbed into the definition of $\xoverline{R}$.

\vskip 4pt

The Teukolsky equation must be supplemented with the appropriate boundary conditions at the event horizon, $r = r_+$, and at asymptotic infinity, $r \to \infty$. Near the event horizon, the radial behaviour of $\psi_4$ is~\cite{Teukolsky:1974yv} 
\beq
\hskip -5pt \psi_4 \sim   Y_{\rm in} (r-r_+)^2 + Y_{\rm out} (r-r_+)^{- 2 i P_+}   \, , \quad r \to r_+ \, , \label{eqn:BChorizon}
\eeq
where $Y_{\rm in}$ and $Y_{\rm out}$ are integration constants, and $P_+ \equiv (am -2 r_+ M \omega)/(r_+ - r_-)$.  To impose the ``purely-ingoing" boundary condition at the horizon, we set $Y_{\rm out} = 0$. On the other hand, the radial behaviour of $\psi_4$ at asymptotic infinity scales as~\cite{Teukolsky:1974yv}
\beq
\psi_4 \sim   Z_{\rm in} \hskip 1pt / r^5 + Z_{\rm out} \hskip 1pt e^{2i \omega \left[ r + 2 M \log r \right]} /r    \, , \quad r \to \infty \, , \label{eqn:BCinf}
\eeq
where $Z_{\rm in}$ and $Z_{\rm out}$ are integration constants. Since the tidal perturbation is sourced in the region between $r=r_+$ and $r \to \infty$, only the outgoing waves at asymptotic infinity carry information about the black hole's tidal responses. This condition requires $Z_{\rm in} = 0$. The coefficients $Y_{\rm in}$ and $Z_{\rm out}$ give rise to fluxes of energy and angular momentum that flow into the black hole horizon and out towards asymptotic infinity, respectively~\cite{Teukolsky:1974yv}.

\vskip 4pt

\section{Tidal perturbations \\ and responses}
In a gravitational system, an object responds to its external tidal field by developing induced mass and current moments. In Newtonian gravity, the \textit{linear} response of a non-rotating object to a slowly-varying tidal field is described by~\cite{poisson_will_2014}
\beq
\hskip -3pt \delta Q_{\ell m} = - \mathcal{N}_\ell \hskip 1pt r_0^{2\ell+1} \left[ 2 \hskip 1pt k_{\ell m}  \hskip 1pt \mathcal{E}_{\ell m} - \tau_0  \hskip 1pt \nu_{\ell m} \hskip 1pt \dot{\mathcal{E}}_{\ell m} + \cdots \right] \, , \label{eqn:response}
\eeq
where $\mathcal{N}_\ell \equiv (\ell-2)! /(2\ell-1)!! $, $\delta Q_{\ell m}$ are the induced mass moments, $\mathcal{E}_{\ell m}$ are the electric-type tidal moments~\cite{Zhang1986}, overdot denotes time derivative, and the ellipses represent dynamical tides, which are suppressed by higher powers in time derivatives. The first term in (\ref{eqn:response}) characterizes the tidal deformability of the object in the static limit, $\omega=0$, and the constants $k_{\ell m}$ are the (dimensionless) Love numbers. The second term describes the energy lost from the tidal environment to the object, with the dissipation numbers $\nu_{\ell m}$ encoding the object's viscosity and $\tau_0$ describing the viscosity-induced delay time.

\vskip 4pt

The induced moments (\ref{eqn:response}) contribute to the total gravitational potential at the exterior of the object. In Fourier space, the potential is $U = - M/r + \Delta U$, where~\cite{poisson_will_2014}\footnote{For notational simplicity, we drop the Dirac delta $\delta (\omega=0)$ for the static terms. }
\beq
\begin{aligned}
\hskip -10pt \Delta U  =  \sum_{\ell m} \frac{\mathcal{E}_{\ell m}  \hskip 1pt r^\ell}{(\ell-1)\ell} \Bigg[ 1 + F_{\ell m} (\omega) \left( \frac{r_0}{r} \right)^{2\ell+1} \Bigg] Y_{\ell m}(\theta, \phi)  \, ,
\label{eqn:NewtonianLove}
\end{aligned}
\eeq
$r_0$ is the radius of the object, $Y_{\ell m}$ is the scalar spherical harmonic, and $F_{\ell m}$ is the response function of the non-rotating object~\cite{Goldberger:2004jt, Goldberger:2005cd, Chakrabarti:2013xza}
\beq
F_{\ell m} (\omega) = 2 \hskip 1pt k_{\ell m} + i \omega \tau_0 \hskip 1pt \nu_{\ell m} + \mathcal{O}(\omega^2) \, .\label{eqn:LinearResponse}
\eeq
The real and imaginary parts of $F_{\ell m}$ capture the tidal deformation and dissipation of the object, respectively~\cite{Goldberger:2005cd, Chakrabarti:2013xza}. The radial dependence of (\ref{eqn:NewtonianLove}) consists of two distinct components: \textit{i)} a characteristic ``growing" term $r^\ell$, which arises due to the contribution of the external tidal perturbation to $U$, and \textit{ii)} a ``decaying" term $r^{-\ell-1}$, which is sourced by the object's linear responses. In General Relativity, the tidal field is also characterized by the magnetic-type moments, $\mathcal{B}_{\ell m}$~\cite{Zhang1986}, to which there is no Newtonian analog. In that case, the induced current moments, the corresponding magnetic Love numbers~\cite{Binnington:2009bb, Damour:2009vw}, and the magnetic dissipation numbers can be defined similarly as (\ref{eqn:response}). 

\vskip 4pt

In addition to quantifying the strength of the perturbation, the moments $\mathcal{E}_{\ell m}$ and $\mathcal{B}_{\ell m} $ also define the characteristic scales of the tidal environment~\cite{HartleThorne1985, Zhang1986}. In particular, the radius of curvature of the environment scales as $\mathcal{R} \sim | \mathcal{E}_{\ell m}|^{-\ell} \sim | \mathcal{B}_{\ell m}|^{-\ell}$~\footnote{Strictly speaking, this scaling only applies to tidal environments which have typical variational lengthscales that are comparable to $\mathcal{R}$ and typical velocities that are close to unity~\cite{HartleThorne1985, Zhang1986}. In this work, we make this notational simplification to ease power counting.}, while its typical variational timescale is $\sim \omega^{-1}$, cf. (\ref{eqn:response}) and (\ref{eqn:LinearResponse}). The lengthscale $\mathcal{R}$ restricts the validity of the multipole expansion of the tidal field to the interval $r < \mathcal{R}$ (see more discussion below). For instance, for binary systems in circular orbits, $\mathcal{R}$ and $ \omega^{-1} $ are of the order of the binary separation.

\vskip 4pt

\section{Perturbed spherically \\ symmetric spacetimes}

Before discussing $\psi_4$ for tidally-perturbed black holes, it is instructive to first investigate the same quantity for the perturbed Schwarzschild spacetime.~This is because Birkhoff's theorem~\cite{Birkhoff, Jebsen} implies that the Schwarzschild metric holds for all spherically-symmetric objects, including those with non-vanishing linear responses (\ref{eqn:LinearResponse}). The corresponding Weyl scalar therefore offers us a convenient way of understanding the way in which the constants $k_{\ell m}$ and $\nu_{\ell m}$ appear in $\psi_4$. 

\vskip 4pt

\begin{figure*}[ht]
\centering
\includegraphics[scale=1, trim = 0 12 0 0]{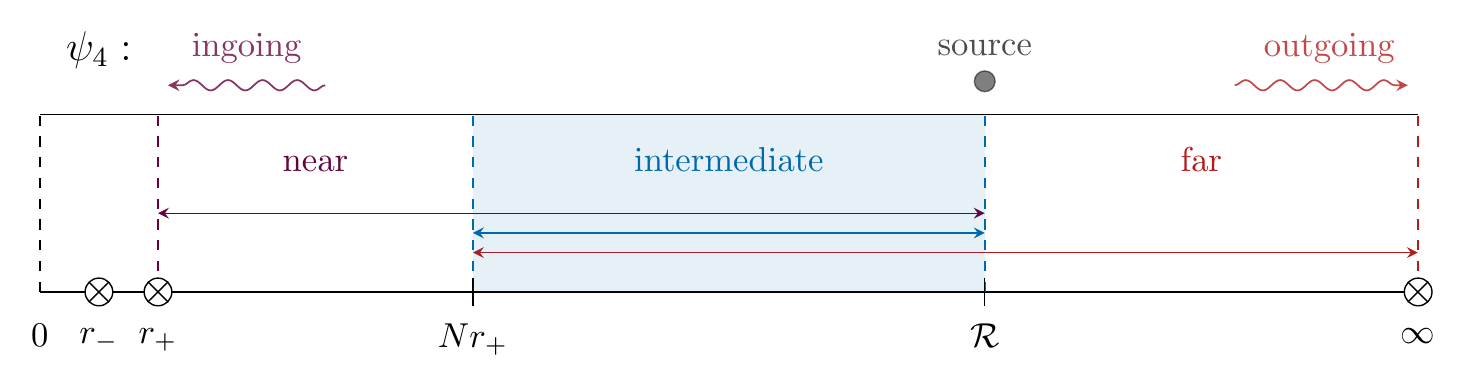}
\caption{Illustration of the different regions of $\psi_4$ for a rotating black hole perturbed by an external tidal field. The near region encodes the purely-ingoing boundary condition at the event horizon $r=r_+$, while the far region captures the outgoing waves at asymptotic infinity $r \to \infty$. The tidal perturbation is sourced in the intermediate region, which captures neither of those conditions. The widths of the regions are determined by two lengthscales: \textit{i)}: the radius of curvature of the tidal environment, $\mathcal{R}$, beyond which the multipole expansion of the tidal field breaks down, and \textit{ii)} the radius $N r_+$, where $N>1$ is a constant value, suitably chosen such that the Kerr background is well described by the Schwarzschild metric at larger distances. }
\label{fig:Regions}
\end{figure*}

Using the perturbed Schwarzschild metric derived in Ref.~\cite{Binnington:2009bb}, we compute the Weyl scalar for a perturbed spherical spacetime, $\psi^{\rm Sph}_4$, by projecting the Weyl tensor onto the Kinnersley null tetrad~\cite{Kinnersley1969}. Since the metric in Ref.~\cite{Binnington:2009bb} was derived in the static limit,  the corresponding Weyl scalar would not encode dissipative effects of the spherical object. Fortunately, this limitation can be circumvented for Schwarzschild black holes, which we shall discuss below. We find that
\beq
\begin{aligned}
 \psi_4^{\rm Sph}(\omega & =0)    =  \sum_{I} \sum_{\ell m} \, \mathcal{M}^I_{\ell m} \,  {}_{-2}Y_{\ell m}(\theta, \phi) \label{eqn:psi4Intermediate}  \\
& \hskip -3pt  \times r^{\ell-2} \left[ G_\ell (r) + 2 \hskip 1pt k^I_{\ell m} \left( \frac{r_0}{r} \right)^{2\ell+1} D_\ell (r)\right] \, , 
\end{aligned}
\eeq
where $I = \{ E, B \}$ labels quantities associated to $\mathcal{E}_{\ell m}$ and $\mathcal{B}_{\ell m}$; the constants $\mathcal{M}_{\ell m}$ scale linearly~\footnote{In general, the perturbed metric also consists of terms involving non-linear couplings between $\mathcal{E}_{\ell m}$ and $\mathcal{B}_{\ell m}$, see e.g. Refs.~\cite{Zhang1986, Poisson:2009qj}. Nevertheless, these non-linearities will be ignored in this work, as we are only interested in the linear response of the object to these tidal moments.} in the tidal moments, and are given by
\beq
 \mathcal{M}_{\ell m}^E = \mathcal{C}_\ell \, \mathcal{E}_{\ell m} \, , \quad \mathcal{M}_{\ell m}^B = i (\ell+1) \hskip 1pt \mathcal{C}_\ell \, \mathcal{B}_{\ell m} / 3 \, , \label{eqn:tidalmoments}
\eeq
with $\mathcal{C}_\ell \equiv 4^{-1} [(\ell+1)(\ell+2)/(\ell-1)\ell]^{1/2}$; and ${}_{-2}Y_{\ell m}$ is the $s=-2$ spin-weighted spherical harmonic~\cite{NP1966, Campbell1971}. From dimensional analysis, we have $\mathcal{M}_{\ell m} r^{\ell-2}\sim r^{-2} (r/\mathcal{R})^{\ell}$: the multipole expansion (\ref{eqn:psi4Intermediate}) is therefore only valid in the $r < \mathcal{R}$ interval, cf. Fig.~\ref{fig:Regions}. The radial functions $G_\ell$ and $D_\ell$ scale as $ 1 + \mathcal{O}( M /r)$ at large distances, $r \gg M$ (see their exact expressions below). By comparing the large-distance limit of (\ref{eqn:psi4Intermediate}) with the Newtonian potential (\ref{eqn:NewtonianLove}), we see that the $r^{\ell-2} G_\ell$ term and the $D_\ell-$dependent term displays the characteristic growing $r^{\ell-2} (1+ \cdots)$ and decaying $r^{-\ell-3} (1+ \cdots)$ behaviour of the tidal and response term, respectively~\footnote{The potential (\ref{eqn:NewtonianLove}) and the Weyl scalar (\ref{eqn:psi4Intermediate}) differ by an overall scaling of $r^{-2}$ because the Weyl tensor consists of two derivatives acting on the metric.}. Crucially, unlike the Newtonian case, $k_{\ell m}$ in (\ref{eqn:psi4Intermediate}) includes both the electric-type and magnetic-type Love numbers. 

\vskip 4pt

\section{Tidal effects of the Schwarzschild black hole}
Although the Schwarzschild solution (\ref{eqn:psi4Intermediate}) cannot be extended towards $r \to r_+$ of a Kerr black hole, it fully describes the near-horizon behaviour of a non-rotating black hole. We therefore discuss how the Love numbers of Schwarzschild black holes~\cite{Binnington:2009bb, Damour:2009vw} are determined from (\ref{eqn:psi4Intermediate}). As we shall see, the analysis presented here is directly applicable to the Kerr black hole below. To this end, it is convenient to introduce the alternative coordinate $y \equiv r/2M - 1$, such that the event horizon is mapped to $y=0$.\footnote{As we elaborate in Appendix~\ref{app:Appendix}, the $y$-coordinate (and the $z$-coordinate which is introduced later for the Kerr black hole) is useful because it allows us to organize the Teukolsky equation in terms of the poles of the differential equation. This organization is especially convenient because it makes clear the terms in the equation that capture the physics of the event horizon. \label{footnote}} The growing term in (\ref{eqn:psi4Intermediate}), expressed in this new coordinate, is 
\beq
\begin{aligned}
r^{\ell-2} \hskip 1pt G_\ell = g_\ell \, y^2 (1+y)^{-2} \hskip 1pt \text{\hyp} (2-\ell, \ell + 3 ; 3; -y) \, , \label{eqn:growFunc}
\end{aligned}
\eeq
where \hyp \hskip 2pt is the hypergeometric function and $g_\ell \equiv \sqrt{\pi} \hskip 2pt 2^{-2\ell-1} (2M)^{\ell-2} \hskip 1pt \Gamma(\ell+3) / \Gamma(\ell+1/2) $. Since the hypergeometric function in (\ref{eqn:growFunc}) is a finite polynomial of degree $\ell-2$~\cite{Bateman}, this solution is proportional to $y^2$  in the $y \to 0$ limit, which is exactly the asymptotic structure of the ingoing wave at the horizon, cf. (\ref{eqn:BChorizon}). On the other hand, the explicit expression for $D_\ell$ is more complicated and has been relegated to Appendix~\ref{app:Appendix}. For our purposes, it suffices to note that $D_\ell$ diverges logarithmically as $y \to 0$. The purely-ingoing boundary condition at the event horizon therefore forces the static response terms in (\ref{eqn:psi4Intermediate}) to vanish identically. This is only possible if the Love numbers of Schwarzschild black holes are all zero~\cite{Binnington:2009bb, Damour:2009vw, Kol:2011vg, Gurlebeck:2015xpa, Hui:2020xxx}.

\vskip 4pt

While (\ref{eqn:psi4Intermediate}) does not shed light on the tidal dissipation of a general spherical object, the dissipative effects of the Schwarzschild black hole can be determined through the Teukolsky equation~\cite{Teukolsky1972a, Teukolsky1973b}, as this equation also describes perturbations with finite frequencies. In this case, we find that the radial part of the Weyl scalar of the perturbed Schwarzschild black hole is 
\beq
\begin{aligned}
\hskip -5pt \psi^{\rm Schw}_{4} \propto  y^2 (1+y)^{-2}  \text{\hyp} (2-\ell, \ell + 3 ; 3 + 2 i \tilde{P}_+ ; -y)  ,  \label{eqn:SchwPsi4omega}
\end{aligned}
\eeq
where $\tilde{P}_+ \equiv - 2 M \omega$ (see Appendix~\ref{app:Appendix} for derivation, where we explain how the $\tilde{P}_+$ term arises from a careful treatment of the Weyl scalar's behaviour at the event horizon). The solution (\ref{eqn:SchwPsi4omega}) clearly reduces to (\ref{eqn:growFunc}) in the static limit. In addition, it encodes the dissipation numbers of Schwarzschild black holes, which can be extracted by expanding (\ref{eqn:SchwPsi4omega}) in the $y \gg 1$ limit. The resulting asymptotic series is (see Appendix~\ref{app:Appendix} for a discussion on the analytic structure of this series)
\beq
\begin{aligned}
 \psi_4^{\rm Schw} \propto & \hskip 5pt y^{\ell-2} \bigg[  \left( 1 + \cdots \right) +  y^{-2\ell-1}  \left(1 + \cdots \right)   \\
& \hskip 0pt  \times\frac{\Gamma(\ell+3) \Gamma(-2\ell-1) \Gamma(\ell+1 + 2 i \tilde{P}_+) }{\Gamma(2\ell+1) \Gamma(-\ell+2) \Gamma(-\ell+ 2 i \tilde{P}_+)}   \bigg] \, , \label{eqn:SchwOmegaCorrections} 
\end{aligned}
\eeq
where ellipses here denote terms that are suppressed by positive powers of $y^{-1}$. Comparing (\ref{eqn:SchwOmegaCorrections}), (\ref{eqn:psi4Intermediate}) and (\ref{eqn:NewtonianLove}), we conclude that the Schwarzschild black hole's response function is given by the coefficient of the decaying terms shown explicitly in (\ref{eqn:SchwOmegaCorrections}), which can be rewritten as
\beq
\begin{aligned}
\hskip -2pt F^{I, \hskip 1pt {\rm Schw}}_{\ell m} & = - i \tilde{P}_+\frac{(\ell-2)! (\ell+2)!}{(2\ell)! (2\ell+1)! } \prod_{j=1}^\ell \Big[ j^2 + 4 \tilde{P}^2_+ \Big]  \, . \label{eqn:SchwResponse}
\end{aligned}
\eeq
The real part of (\ref{eqn:SchwResponse}) is zero -- as discussed above, the Love numbers of non-rotating black holes vanish identically. Substituting $\tilde{P}_+ = - 2 M \omega$ into (\ref{eqn:SchwResponse}), we find that the response function is proportional to $ i \omega (2M) $, which is exactly the structure of the imaginary part of (\ref{eqn:LinearResponse}), with $\tau_0 = 2M$ being the black hole light crossing time. The imaginary part of (\ref{eqn:SchwResponse}) therefore represents the dissipation numbers of the Schwarzschild black hole. This expression correctly reproduces known results in the literature for the few low-order dissipation numbers (note our different choice of normalization in (\ref{eqn:response}))~\cite{Goldberger:2005cd, Poisson:2004cw, Chakrabarti:2013lua}.

\vskip 4pt

\section{Tidal effects of \\ the Kerr black hole}
The solution (\ref{eqn:SchwPsi4omega}) is only valid in the ``intermediate region" of a tidally-perturbed rotating black hole, which we define as the interval $N r_+ < r < \mathcal{R}$, where $N$ is a constant greater than order unity, cf. Fig.~\ref{fig:Regions}. This is because (\ref{eqn:SchwPsi4omega}) does not capture the boundary conditions at $r=r_+$ and $r \to \infty$. To parameterize the full radial behaviour of $\psi_4$, we introduce the function, $R$, through the ansatz 
\beq
\begin{aligned}
\hskip -4pt \rho^4 \psi_4 = & \sum_{I}  \sum_{\ell m} \, \mathcal{M}^I_{\ell m}  e^{- i \omega v + i m \phi} \hskip 1pt   R^I_{\ell m} (r)  {}_{-2}S_{\ell m}(\theta)  \, . \label{eqn:separabletidal}
\end{aligned}
\eeq
In the adiabatic limit, $\omega \to 0$, the angular function $  {}_{-2} S_{\ell m} e^{i m \phi} = {}_{-2} Y_{\ell m} + \mathcal{O}(a \omega) $ is well approximated by the spin-weighted spherical harmonic. Comparing the ansatz (\ref{eqn:separabletidal}) with (\ref{eqn:psi4Intermediate}), we see that the former contains an additional dependence on $\theta$ through the $\rho^4$ factor, which becomes purely radial in the Schwarzschild limit. This factor is included in our ansatz because we want (\ref{eqn:separabletidal}) to mimic the separable form (\ref{eqn:mode}) as closely as possible. In fact, they are identical up to the redefinition $\mathcal{M} R \to \xoverline{R}$: the tidal moments only affect the amplitudes of the perturbation modes. The identification between (\ref{eqn:separabletidal}) and (\ref{eqn:mode}) is important because, as we emphasized above, the separable form (\ref{eqn:mode}) applies to any form of linear gravitational perturbation~\cite{Teukolsky1972a, Teukolsky1973b}. Our ansatz must therefore be true for a rotating black hole immersed in the tidal environment, with $R$ satisfying the Teukolsky equation. 

\vskip 4pt

The Love numbers and dissipation numbers of the Kerr black hole are determined through the solution of $\psi_4$ in the ``near region", which is the interval $r_+ < r < \mathcal{R}$ that encapsulates the boundary condition at the event horizon, cf. Fig.~\ref{fig:Regions}. Introducing the $z \equiv (r-r_+)/(r_+ - r_-)$ coordinate, we find that the radial part of the near-zone $\psi_4$ is (see Footnote~\ref{footnote} and Appendix~\ref{app:Appendix}) 
\beq
\begin{aligned}
\psi_{4} & \propto \left[ z \hskip 1pt (r_+ - r_-) + r_+ - i a \cos \theta \right]^{-4} \\[2pt]
& \times   z^2 (1+z)^2  \,  \text{\hyp} (2-\ell, \ell + 3  ; 3 + 2 i P_+; -z) \label{eqn:LOnear} \, .
\end{aligned}
\eeq 
This solution is the rotational generalization of its Schwarzschild analog (\ref{eqn:SchwPsi4omega}). This can be seen most transparently by setting $a=0$, in which case $P_+=\tilde{P}_+$, the first line of (\ref{eqn:LOnear}) becomes proportional to $(1+z)^{-4}$, and the $y-$ and $z-$coordinates coincide. The response function of the Kerr black hole can therefore be derived by following the same prescription as above, to which we obtain\footnote{Our result (\ref{eqn:KerrResponse}) generalizes Eq. (23) of Ref.~\cite{LeTiec:2020spy} to non-vanishing values of $\omega$. The apparent difference between the functional form of our result and theirs is resolved by using the identitiy $\Gamma(n+k)\Gamma(n-k) = [\pi k / \sin (\pi k )] \prod_{j=1}^{n-1} (j^2 - k^2)$. Crucially, while they interpreted their findings as Love numbers, we show that the non-vanishing component in (\ref{eqn:KerrResponse}) are instead the dissipation numbers of Kerr black holes. }
\beq
\begin{aligned}
F^{I, {\rm Kerr}}_{\ell m} & = - i P_+ \frac{(\ell-2)! (\ell+2)!}{(2\ell)! (2\ell+1)! } \prod_{j=1}^\ell \left[ j^2 + 4 P^2_+ \right]  \, . \label{eqn:KerrResponse}
\end{aligned}
\eeq
Remarkably, the real part of (\ref{eqn:KerrResponse}) vanishes identically: the Love numbers of rotating black holes are \textit{zero}. On the other hand, the dissipation numbers in (\ref{eqn:KerrResponse}) are non-trivial rotational generalizations of those in (\ref{eqn:SchwResponse}). For example, unlike the Schwarzschild case, the $P_+ = (a m - 2 r_+ M \omega)/(r_+ - r_-)$ factor in (\ref{eqn:KerrResponse}) implies that the Kerr black hole would experience tidal dissipation even at $\omega=0$, as the black hole's rotational motion sources the relative time dependence with the static environment~\footnote{In other words, in the rest frame/co-rotating frame of the Kerr black hole, a static tidal field would be perceived to rotate at a frequency that is proportional to $a$ and therefore induces tidal dissipation, cf. (\ref{eqn:response}).}. In addition, $P_+$ is not directly proportional to $\omega$, which scales as three powers in velocities for binary systems: tidal dissipation of rotating black holes in this case is thus enhanced by 1.5PN orders compared to that of non-rotating black holes~\cite{Poisson:1994yf, Tagoshi:1997jy}. Furthermore, $P_+$ can be either positive or negative: the response function (\ref{eqn:KerrResponse}) encodes the superradiance phenomenon that is characteristic of rotating black holes~\cite{Zeldovich:1972spj, Press:1972zz, Starobinsky:1974spj, Page:1976df, Detweiler:1980uk, Baumann:2019eav}. Importantly, our result (\ref{eqn:KerrResponse}) is true for all values of black hole spin, $0 \leq a \leq M$; for both the electric-type and magnetic-type perturbations, $I = \{E, B\}$; and for all tidal moments in the multipole expansion, $\ell \geq 2$, $ |m| \leq \ell$.

\section{From tidal effects to gravitational-wave fluxes}
The near-zone solution (\ref{eqn:LOnear}) is limited to the $r < \mathcal{R}$ interval because the multipole expansion breaks down at larger distances. To map the black hole's responses (\ref{eqn:KerrResponse}) onto gravitational-wave observables, a matched asymptotic expansion~\cite{Starobinsky:1974spj, Page:1976df, Detweiler:1980uk, Baumann:2019eav, bender1999advanced, holmes1998introduction}  between the near region and the ``far region" must be performed. The latter region lies in the interval $N r_+ < r < \infty$, which includes the source of perturbation and the outgoing boundary condition at asymptotic infinity, cf. Fig.~\ref{fig:Regions}. Through these matching computations, one can calculate various physical quantities, including the gravitational-wave fluxes propagating towards $r \to r_+$ and $r \to \infty$~\cite{Teukolsky:1974yv}. In fact, for extreme mass-ratio inspirals, these fluxes have been computed to remarkable high orders in $\omega$~\cite{Mano:1996vt, Mano:1996gn, Mino:1997bx, Hughes:2001jr, Fujita:2014eta, Sasaki:2003xr}. In that case, the Love numbers leave their imprints on the flux emitted to asymptotic infinity~\cite{Flanagan:2007ix, Vines:2011ud}. This is because the tidal deformations of the binary components would source additional mass and current moments, with the binaries' orbital motions inducing the necessary time dependences for these moments to emit additional fluxes to asymptotic infinity. On the other hand, the dissipation numbers of black holes describe the absorptive nature of the event horizon, and are thus encoded in the flux flowing into the horizon~\cite{Poisson:1994yf, Tagoshi:1997jy}. These fluxes would also backreact on the orbital motion, thereby affecting the binary's conservative dynamics in a non-trivial manner~\cite{Mano:1996vt, Mano:1996gn, Mino:1997bx, Hughes:2001jr, Fujita:2014eta, Sasaki:2003xr}. In any case, since these fluxes are physically measurable quantities, the Love numbers and dissipation numbers (\ref{eqn:KerrResponse}) that are encoded in them have unambiguous physical interpretations on observable waveforms.

\vskip 4pt 

\section{Love Numbers are Unambiguous}
Our discussion above also offers a resolution to the ambiguity issue of Love numbers in General Relativity~\cite{Gralla:2017djj}. In Refs.~\cite{Fang:2005qq, Pani:2015hfa, Gralla:2017djj}, it was described how a class of coordinate transformations, denoted here schematically by $r \to \bar{r}$, would mix the tidal and response terms in the perturbed metric of a generic object. Those transformations would lead to different coefficients between the $ r^{-\ell-1} (1 + \cdots)$ and $ \bar{r}^{\hskip 1pt -\ell-1} (1 + \cdots) $ decaying terms, thereby apparently changing the values of the Love numbers. However, those studies only focused on the near region but did not consider the far region~\footnote{To the best of our knowledge, the far region has been largely ignored in the tidal deformability literature. This is the case presumably because $\omega$ is often set to zero at the onset of those calculations. Such a procedure, while natural for computing the static response, would inadvertently conceal the true boundary condition at $r \to \infty$. In particular, taking the adiabatic limit, with the $r-$coordinate held fixed, only formally extends the regime of validity of the near-zone solution to $\mathcal{R} \sim \omega^{-1} \to \infty$, cf. Fig.~\ref{fig:Regions}. Despite its enlarged radial interval, this near-zone solution can never capture the only-outgoing boundary condition at asymptotic infinity (\ref{eqn:BCinf}). Instead, this boundary condition can only be described by the far-zone solution, which is obtained by rescaling the radial Teukolsky equation (see Appendix~\ref{app:Appendix}) with the $x \equiv \omega r$ coordinate, and keeping $x$ fixed when solving for the equation in the adiabatic limit~\cite{Starobinsky:1974spj, Page:1976df, Detweiler:1980uk, Baumann:2019eav}.}. As a result, the outgoing flux and the boundary condition at asymptotic infinity were not taken into account. This is a subtle but important shortcoming, because a coordinate transformation which is only performed in the near region, but not consistently over the entire spacetime, implicitly alters the asymptotic behaviour of the outgoing waves at $r \to \infty$. Furthermore, a coefficient that appears before any decaying term has no real physical interpretation unless it is matched with observables. To resolve the ambiguity issue in those studies, one must find the metric in the far region, perform the coordinate transformation $r \to \bar{r}$ accordingly, and match the far-zone solution with the near-zone solution. By virtue of general covariance, the inferred values of the Love numbers must remain unaltered. In short, we argue that Love numbers in General Relativity can be unambiguously defined when \textit{both} the boundary conditions that characterize the nature of the object and that of outgoing waves at asymptotic infinity are considered.

\vskip 4pt

\section{Conclusions}
To summarize, we showed that the Love numbers of Kerr black holes are all zero. We also derived the dissipation numbers of Kerr black holes, which are shown in (\ref{eqn:KerrResponse}) for our choice of normalization (\ref{eqn:response}). In addition, we addressed the ambiguity issue of Love numbers in General Relativity, which we argued is resolved when both the boundary conditions of the perturbed object and that of outgoing waves at asymptotic infinity are taken into account. The imprints of the Kerr black holes' tidal effects on observables have, in fact, already been computed in the literature~\cite{Mano:1996vt, Mano:1996gn, Mino:1997bx, Hughes:2001jr, Fujita:2014eta, Sasaki:2003xr}. Our results are therefore free of coordinate ambiguity and have robust physical interpretations on gravitational waveforms. Finally, we note that the Teukolsky equation does not describe any metric perturbation of Petrov type II class, i.e. a gravitational perturbation that involves a non-linear combination of transverse, longitudal, and coulombic contributions~\cite{Szekeres:1965ux}.  In future work, we hope to explore if the conclusions drawn in this work would still hold for this more general type of perturbation.

\begin{acknowledgments}

I am grateful to Marc Casals, Tanja Hinderer, Austin Joyce, Alexandre Le Tiec,  Rafael Porto, and John Stout for stimulating discussions. I thank the University of Amsterdam for its extended hospitality in times of restricted travels. This work is supported by the Netherlands Organisation for Scientific Research (NWO) through a Rubicon Fellowship. 

\end{acknowledgments}

\appendix

\section{} \label{app:Appendix}

\subsection{The $\psi_4$ Weyl scalar and the null tetrad}

The $\psi_4$ Weyl scalar of the Newman-Penrose formalism~\cite{NP1962} encapsulates the dominant behaviour of the outgoing gravitational waves at asymptotic infinity. Denoting the set of null tetrad vectors by $\{l^\mu, n^\mu, m^\mu, \bar{m}^\mu \}$, where $l^\mu$ and $n^\mu$ are real, $m^\mu$ is complex and $\bar{m}^\mu$ is its conjugate, $\psi_4$ is defined as
\begin{equation}
\psi_4 \equiv  C_{\mu \nu \rho \sigma} n^\mu \bar{m}^\nu n^\rho \bar{m}^\sigma \, , \label{eqn:psi4def}
\end{equation}
where $C_{\mu \nu \rho \sigma}$ is the Weyl tensor. In this work, we adopt the Kinnersley null tetrad~\cite{Kinnersley1969}. In the ingoing-Kerr coordinates (\ref{eqn:Kerrcoord}), these tetrad vectors are
\begin{equation}
\begin{aligned}
l^\mu & =  \left[ 2(r^2 + a^2)/\Delta, 1, 0, 2a/\Delta \right] \, , \\
n^\mu & = \left[ 0, - \Delta/(2\Sigma), 0, 0 \right] \, , \\
m^\mu & = \left[ i a \sin \theta , 0, 1, i /\sin \theta \right]/[ \sqrt{2} \hskip 1pt (r + i a \cos \theta) ] \, , \label{eqn:Ktetrad}
\end{aligned}
\end{equation}
which satisfy the orthonormality relation $l^\mu n_\mu = - m^\mu \bar{m}_\mu= -1$, while other inner products vanish. 

\subsection{The $\psi_4$ Weyl scalar of a perturbed spherical object}

The quantity $\psi_4^{\rm Sph}$ in (\ref{eqn:psi4Intermediate}) is obtained through an explicit computation of (\ref{eqn:psi4def}). In this case, $C_{\mu \nu \rho \sigma}$ is the Weyl tensor of the perturbed Schwarzschild metric derived in Ref.~\cite{Binnington:2009bb}, and $a$ is set to zero in the tetrad (\ref{eqn:Ktetrad}). While the growing function $r^{\ell-2} G_{\ell}$ can be expressed succinctly as (\ref{eqn:growFunc}), the decaying function $D_\ell$ is more complicated. In particular, in the $y = r/2M-1$ coordinate,
\begin{equation}
\begin{aligned}
D_\ell = y^2  \left[ d_\ell '' - \frac{2 \ell }{1+y}  d_\ell ' + \frac{\ell (\ell + 1)}{(1+y)^2} d_\ell  \right] \, , \label{eqn:decayFunc}
\end{aligned}
\end{equation}
where prime denotes derivative with respect to $y$ and $d_\ell = 2 \hskip 1pt [\ell (\ell+1) (\ell+2) ]^{-1} \big[ \hskip 1pt \text{\hyp} \left( \ell, \ell+1; 2\ell+2; (1+y)^{-1} \right) + (\ell/2) \hskip 1pt  \text{\hyp} \left( \ell+1, \ell+1; 2\ell+2; (1+y)^{-1} \right) \big]$. Since the first two arguments of the hypergeometric functions here are positive integers, (\ref{eqn:decayFunc}) diverges logarithmically in the $y \to 0$ limit~\cite{Bateman}. 

\subsection{The Teukolsky equation}

As described in the main text, the radial function $R$ in (\ref{eqn:separabletidal}) satisfies the Teukolsky equation. In the Kinnersley null tetrad (\ref{eqn:Ktetrad}), this equation reads 
\begin{equation}
\begin{aligned}
& \frac{\d^2 R}{\d r^2} +  \left( \frac{2 i P_+ - 1}{r-r_+} - \frac{2 i P_- + 1}{r - r_-} - 2 i \omega \right)\frac{\d R}{\d r} \\
& + \bigg( \hskip 2pt \frac{4 i P_-}{(r-r_-)^2} - \frac{4 i P_+}{(r-r_+)^2} +  \frac{ A_- + i B_-}{(r-r_-) (r_+ - r_-)}  \\
&  - \frac{ A_+ + i B_+}{(r-r_+) (r_+ - r_-)} \hskip 2pt \bigg) R = \frac{T}{\Delta} \, , \label{eqn:radialeqn}
\end{aligned}
\end{equation}
where $P_{\pm}, A_{\pm}$ and $B_{\pm}$ are
\begin{equation}
\begin{aligned}
 P_\pm & \equiv \frac{a m - 2 r_\pm M \omega }{r_+ - r_-} \, , \qquad B_\pm  \equiv 2 r_\pm \omega \, ,  \\
A_{\pm} & \equiv  E_{\ell m} -2 - 2 (r_+ - r_-) P_\pm \omega \\
& \hskip 10pt - ( r_\pm + 2M  ) r_\pm \omega^2  \, , 
\end{aligned}
\end{equation}
$E_{\ell m} = \ell(\ell+1) + \mathcal{O}(a \omega)$ is the angular eigenvalue and $T$ represents a generic source of energy momentum~\cite{Teukolsky1972a, Teukolsky1973b} (in Ref.~\cite{Teukolsky:1974yv}, the $-2(2s+1) i \omega r$ factor in the Teukolsky equation should instead read $2(2s-1)i \omega r$. This error has also been pointed out in e.g. Ref.~\cite{Chatziioannou:2012gq}). We have casted (\ref{eqn:radialeqn}) in a form whereby the poles of the differential equation are clearly identifiable, cf. Fig.~\ref{fig:Regions}. Since analytic solutions of (\ref{eqn:radialeqn}) cannot simultaneously capture both of the boundary conditions at $r=r_+$ and $r \to \infty$, approximate solutions in the near and far regions have to be separately constructed. These solutions are obtained by leveraging on the adiabatic limit, in which (\ref{eqn:radialeqn}) is expanded perturbatively in the dimensionless parameter $M \omega \ll 1$~\cite{Starobinsky:1974spj, Page:1976df}. From the virial theorem, we conclude that the $M\omega-$expansion is equivalent to an expansion in the strength of the gravitational source $T$. For instance, for a binary system, $T /\Delta \sim \mu / r  \sim (M\omega)^{2/3} \hskip 1pt q \hskip 1pt  (1+q)^{-1/3}$~\cite{Mino:1997bx}, where $\mu$ is the companion's mass and $q \equiv \mu/M$ is the mass ratio.

\subsection{Derivation of the near-zone solution}

To solve for (\ref{eqn:radialeqn}) in the near region, we introduce the coordinate $z \equiv (r-r_+)/(r_+ - r_-)$, such that the event horizon is mapped to $z=0$. We then expand (\ref{eqn:radialeqn}) at leading order in $M\omega$ while keeping $z$ fixed, in order to obtain the near-zone radial equation. The virial theorem implies that we can simultaneously take $T = 0$ at this order. Importantly, since the $P_+$ factor appears as the coefficient of the $(r-r_+)^{-2}$ pole in (\ref{eqn:radialeqn}), it dominates the behaviour of the solution at the event horizon.
 It is therefore critical that $P_+$ is held fixed in the $M\omega$-expansion, such that all of the physics associated to the event horizon are preserved in our perturbative expansion (see Footnote~9 of~Ref.~\cite{Baumann:2019eav} for the same comment, though in a slightly different context). All in all, we find that the near-zone radial equation at leading order is
\begin{equation}
\begin{aligned}
& \frac{\d^2 R}{\d z^2} +  \left( \frac{2 i P_+-1}{z} - \frac{2 i P_+ + 1}{z+1} \right) \frac{\d R}{\d z} \\
& + \left( \frac{4 i P_+}{(z+1)^2}- \frac{4 i P_+}{z^2}  - \frac{\ell(\ell+1) -2}{z(1+z)} \right) R = 0 \, .
   \end{aligned}
\end{equation}
Imposing the purely-ingoing boundary condition at the horizon, and restoring the factor $\rho^{-4} = [z \hskip 1pt (r_+ - r_-) + r_+ - i a \cos \theta]^{-4}$, we obtain the near-zone solution of $\psi_4$ in (\ref{eqn:LOnear}). This derivation trivially applies to the Schwarzschild solution (\ref{eqn:SchwPsi4omega}). Crucially, from the linear responses (\ref{eqn:SchwResponse}) and (\ref{eqn:KerrResponse}), we see that the black hole's tidal dissipation is precisely encoded in $P_+$ (which we would have been blind to for the Schwarzschild black hole had we set $\omega=0$ in our perturbative expansion).

\subsection{The large-distance limit of the \\ near-zone solution}

\textit{The large-distance limit of the near-zone solution.---}The response function of the Kerr black hole is inferred from the decaying terms in the asymptotic series of the near-zone solution, cf. (\ref{eqn:SchwOmegaCorrections}) for its Schwarzschild analog. \textit{A priori}, it is not obvious that the decaying terms should be present in the series. This is because the hypergeometric function $ \text{\hyp} (2-\ell, \ell + 3  ; 3 + 2 i P_+; -z) $ in (\ref{eqn:LOnear}), whose first argument is a non-negative integer, is a polynomial of degree $\ell-2$~\cite{Bateman}. The near region would therefore naively seem to only consist of growing terms but not decaying terms. 

However, it is known in the literature~\cite{Starobinsky:1974spj, Page:1976df, Detweiler:1980uk, Baumann:2019eav} that, in order to perform a matched asymptotic expansion between the near and far regions, one must treat $\ell$ as a general real value in the asymptotic expansion before specializing it to an integer, as we did in (\ref{eqn:SchwOmegaCorrections}). The result of this procedure is the presence of finite decaying terms, which are necessary for completing the matching computation. One may view the analytic continuation in $\ell$ as a way of exploiting the analytic properties of the hypergeometric function, in order to construct two linearly-independent basis functions over the entire radial interval~\cite{Mano:1996vt, Mano:1996gn, Mino:1997bx}. In fact, this concept can be formalized through the introduction of the so-called ``renormalized angular momentum number"~\cite{Mano:1996vt, Mano:1996gn, Mino:1997bx}, a key quantity that has allowed for accurate matching computations at high perturbative orders. Ultimately, the validity of the series (\ref{eqn:SchwOmegaCorrections}), and its Kerr generalization, is justified by its ability to capture all of the physics in the far region through the matched asymptotic expansion.

\newpage

\bibliographystyle{apsrev4-1}
\bibliography{KerrLove}

\end{document}